\documentclass[english]{sig-alternate-05-2015}

\begin{document}

\setcopyright{acmcopyright}





%

\title{Semantic Scan: Detecting Subtle, Spatially Localized Events in Text Streams
}
%
%
%
%
%

\numberofauthors{6} 
%
\author{
%
%
\alignauthor
Abhinav Maurya\titlenote{Work performed while at Event and Pattern Detection Laboratory, Carnegie Mellon University.}\\
       \affaddr{H. J. Heinz III College}\\
       \affaddr{Carnegie Mellon University}\\
       \affaddr{Pittsburgh, PA 15213}\\
       \email{ahmaurya@cmu.edu}
\alignauthor 
Kenton Murray\raisebox{10pt}{$\ast$}\\
      \affaddr{Computer Science \& Engineering}\\
      \affaddr{University of Notre Dame}\\
      \affaddr{Notre Dame, IN 46556}\\
      \email{kmurray4@nd.edu}
\alignauthor
Yandong Liu\raisebox{10pt}{$\ast$}\\
      \affaddr{Uber Inc}\\
      \affaddr{San Francisco, CA 94103}\\
      \email{yandongl@cs.cmu.edu}
\and
\alignauthor
Chris Dyer \\
      \affaddr{Language Technologies Institute}\\
      \affaddr{Carnegie Mellon University}\\
      \affaddr{Pittsburgh, PA 15213}\\
      \email{cdyer@cs.cmu.edu}
\alignauthor
William W. Cohen \\
      \affaddr{Machine Learning Department}\\
      \affaddr{Carnegie Mellon University}\\
      \affaddr{Pittsburgh, PA 15213}\\
      \email{wcohen@cs.cmu.edu}
\alignauthor
Daniel B. Neill\titlenote{Corresponding Author}\raisebox{10pt}{$\ast$} \\
      \affaddr{H. J. Heinz III College}\\
      \affaddr{Carnegie Mellon University}\\
      \affaddr{Pittsburgh, PA 15213}\\
      \email{neill@cs.cmu.edu}
}

\maketitle
\begin{abstract}
Early detection and precise characterization of emerging topics in text streams can be highly useful in applications such as timely and targeted public health interventions and discovering evolving regional business trends. Many methods have been proposed for detecting emerging events in text streams using topic modeling. However, these methods have numerous shortcomings that make them unsuitable for rapid detection of locally emerging events on massive text streams. In this paper, we describe Semantic Scan (SS) that has been developed specifically to overcome these shortcomings in detecting new spatially compact events in text streams.

Semantic Scan integrates novel contrastive topic modeling with online document assignment and principled likelihood ratio-based spatial scanning to identify emerging events with unexpected patterns of keywords hidden in text streams. This enables more timely and accurate detection and characterization of anomalous, spatially localized emerging events. Semantic Scan does not require manual intervention or labeled training data, and is robust to noise in real-world text data since it identifies anomalous text patterns that occur in a cluster of new documents rather than an anomaly in a single new document.

We compare Semantic Scan to alternative state-of-the-art methods such as Topics over Time, Online LDA, and Labeled LDA on two real-world tasks: (i) a disease surveillance task monitoring free-text Emergency Department chief complaints in Allegheny County, and (ii) an emerging business trend detection task based on Yelp reviews. On both tasks, we find that Semantic Scan provides significantly better event detection and characterization accuracy than competing approaches, while providing up to an order of magnitude speedup.
\end{abstract}

%
%



%
%

%
%





\section{Introduction}

Text streams are ubiquitous in data processing and knowledge discovery workflows. Their analysis and summarization is difficult because of their unstructured nature, sparsity of the canonical bag-of-words representation, the massive scale of web-scale text streams like Twitter and Yelp Reviews, and the noise present due to word variations from misspellings, dialects, and slang.

Topic modeling is a mixed-membership model used to summarize a corpus of text documents using a set of latent topics, where each topic is a sparse distribution on words. However, traditional topic modeling methods like Latent Dirichlet Allocation (LDA) are too slow for analyzing web-scale text streams, and also assume that there is no concept drift in the topics being learned over time. Variations like \emph{Online LDA} \cite{hoffman2010online}, \emph{Dynamic Topic Models} \cite{dDTM}, and \emph{Topics over Time} \cite{wang2006topics} relax the assumption that there is no concept drift in the learned topics with time, but make strong assumptions about the smooth evolution of topics with time, making them poor models of the ``bursty'' dynamics that are often observed as new topics appear \cite{kleinberg:2002}.


In this paper, we propose \emph{Semantic Scan (SS)} which was developed to overcome these shortcomings in the scalable detection of spatially localized emerging topics in text streams. SS is a novel framework for detecting anomalous patterns in spatio-temporal free text data. It integrates contrastive topic modeling with online document assignment and principled likelihood ratio-based spatial scanning to identify emerging events with unexpected patterns of keywords hidden in text streams, thus enabling more timely and accurate detection and characterization of anomalous, spatially compact outbreaks. There is no need for manual intervention, labeled training data, or expensive data preprocessing, as SS robustly handles noisy data complete with uncommon phrases, misspellings, and incorrect entries.

We compare our method with three state-of-the-art methods - \emph{Topics over Time} \cite{wang2006topics}, \emph{Online LDA} \cite{hoffman2010online}, and \emph{Labeled LDA} \cite{ramage2009labeled} - to prove the effectiveness and efficiency of our methods. We carry out the comparison on two real-world detection tasks. The first is a disease surveillance task based on monitoring free-text Emergency Department chief complaints in Allegheny County. The second is the task of detecting emerging business trends based on Yelp reviews. On both tasks, SS enables significant improvement in detection time and the percentage of events detected while incurring a fraction of running time compared to competing methods.

Stakeholders using an algorithm need not only the ability to detect an event but also precisely characterize the nature and scope of the event. We test the ability of SS to characterize the emerging event using three metrics: (i) Hellinger distance between the detected emerging topic and the empirical distribution of words in the documents known to contain the emerging topic, (ii) Spatial Overlap - the Jaccard similarity between the set of truly affected locations and the set of locations that were detected to be affected, and (iii) Document Overlap - the Jaccard similarity between the set of documents containing the true emerging topic and the set of documents that were detected to contain the detected emerging topic. We find that SS performs better than competing approaches in event characterization, and therefore serves as a useful tool in the event detection and characterization toolkit.

\section{Background and Related Work}
Both spatial scan statistics and topic models are popular machine learning methods for identifying structure in data. Spatial 
scan aims to discern anomalous patterns within spatially and temporally located data, determining if portions of the dataset 
cannot be explained by an underlying, baseline process and therefore may be of potential interest. Topic models attempt to 
discover latent mixtures of topics (where each topic is a probability distribution over words) that describe a corpus of 
unstructured text data.  While extensions to topic models have incorporated spatial and temporal information, no work has been 
done to detect anomalous spatio-temporal regions using topic models. Likewise, spatial scan statistics have been extended to a variety of 
data types, but have not previously been able to deal with unstructured free text data. Finally, as compared to previous dynamic and online LDA approaches, which assume that topics evolve smoothly over time, our contrastive topic modeling approach is much more effective for detecting newly emerging topics and their corresponding, spatially localized events.

\subsection{Spatial Scan Statistics}

The spatial scan statistic~\cite{kulldorff} is a powerful method for spatial event 
surveillance which detects anomalous spatial or space-time clusters that are not well explained by a baseline process. An 
extension of scan statistics~\cite{naus}, which attempt to determine if a point process is random, spatial scan statistics 
are frequently used by the public health community for detecting spatial clusters of diseases such as breast 
cancer~\cite{breast}, leukemia~\cite{lukemia}, and West Nile virus~\cite{nile}. They have also been broadly applied to other 
structured space-time data in tasks such as crime detection~\cite{crimer}.

Kulldorff's spatial scan~\cite{kulldorff,kulldorff01} searches over geographic areas consisting of circles of varying radii 
centered at each of the monitored spatial locations.  Over this set of regions, it maximizes a likelihood ratio statistic that 
compares the observed count to the expected count i.e. the baseline, where baselines are estimated from population data or from time 
series analysis of historical data~\cite{ebp}. This approach has been extended to other spatial areas such as 
rectangles~\cite{rectangles}, ellipses~\cite{ellipses}, and irregularly shaped regions~\cite{irregularD,irregularP,irregularT}, 
and has been generalized to ``subset scan'', which enables the efficient identification of anomalous subsets in more general 
datasets with spatial, temporal, or graph constraints~\cite{LTSS}.

Spatial scan statistics typically monitor a set of known spatial locations, $\{s_1 \ldots s_N\}$, where each location $s_i$ has
a time series of observed counts $c_i^t$ and a time series of expected counts $b_i^t$.  They scan over the set of space-time regions $S$
consisting of spatially constrained subsets of locations $S_{spatial} \subseteq \{s_1 \ldots s_N\}$ for time durations
$W \in \{1 \ldots W_{max}\}$, and maximize a likelihood ratio statistic $F(S)$, where:
\begin{equation}\label{scanstatisticequation}
F(S) = \log{\frac{\mbox{Pr}(\mbox{Data}\:|\:H_1(S))}{\mbox{Pr}(\mbox{Data}\:|\:H_0)}}
\end{equation}

In this equation, the alternative hypothesis $H_1(S)$ typically assumes a multiplicative increase in counts $c_i^t$ for the 
given space-time region $S$, while the null hypothesis assumes that all counts $c_i^t$ are generated from some 
distribution with means proportional to $b_i^t$.  Here we use the expectation-based Poisson scan statistic~\cite{ebp}, commonly used to 
model count data, which assumes that $c_i^t \sim \mbox{Poisson}(b_i^t)$ under $H_0$, while under $H_1(S)$, we have $c_i^t \sim 
\mbox{Poisson}(qb_i^t)$ for $s_i \in S$ for some constant $q > 1$.  Assuming the maximum likelihood estimate of $q$, the log-likelihood ratio simplifies to:
\begin{equation}\label{ebpequation}
F(S) = \begin{cases}       
C \log{\frac{C}{B}} + B - C; & C > B \\
0; & C \leq B
\end{cases}
\end{equation}
where $C$ and $B$ are respectively the aggregate count $\sum c_i^t$ and aggregate baseline $\sum b_i^t$ for space-time region 
$S$.  While this formulation focuses on detecting regions with higher than expected counts, many other variants exist for identifying 
decreased counts, higher counts inside the region versus outside~\cite{kulldorff}, or incorporating other parametric models such 
as Gaussian or exponential counts~\cite{LTSS}.

\subsection{Topic Modeling}

Topic modeling is a popular set of methods for dealing with unstructured data and free text. In general, topic modeling 
algorithms attempt to fit a latent mixture of thematic topics to each individual document in a corpus. Each topic is a 
distribution over words in the corpus, and each document is represented as a mixture of these topics. Given a corpus of 
documents with only observed words, topic modeling algorithms attempt to learn the posterior distribution over words for each 
topic and over topics for each document.  One of the most well-known topic models, Latent Dirichlet Allocation (LDA), has 
become commonly used for unsupervised text corpus modeling~\cite{blei2003latent}. Topic models enhance text classification by 
allowing multiple topics to exist within a document, and by allowing words to have a probabilistically assigned likelihood of 
being generated from a specific topic.

LDA models a corpus of documents $d=1\ldots D$, each with a potentially different number of words $N_d$, from a vocabulary $V$. 
The model assumes a generative process for a corpus where each document $d$ has a mixture of topics, represented as a 
multinomial distribution $\theta_d$. Each word $i$ in the document has an individual topic assignment $z_{di} = k$, and then a 
word $w_{di}$ is drawn from the vocabulary $V$ using the multinomial distribution over words $\phi_k$ for the selected topic.

Attempting to modify LDA to account for topic shift over time is an active area of research. Dynamic Topic Models~\cite{dDTM} 
allow the Dirichlet hyperparameters $\alpha$ and $\beta$, for $\theta$ and $\phi$ respectively, to vary over time using a Markov 
assumption with Gaussian noise. This allows the topics at each time $t$ to ``smoothly'' evolve from the previous topics at time 
$t-1$.  Continuous time dynamic topic models~\cite{cDTM} remove the assumption of discrete time steps by using Brownian motion. 
The evolution of the hyperparameters allows the same topics to evolve over time; this differs from our approach described below, as we do not let topics 
evolve, but instead identify newly emerging topics.  In contrast to the previous methods, Topics Over Time~\cite{wang2006topics} is a method where the topics 
are fixed but the topics' relative occurrences and correlations change over time.  Similarly, \cite{dynamicmixture} modifies the 
basic LDA model to allow topic mixtures, $\theta$, to vary over time according to a Markov assumption, but keeps $\phi$, 
$\alpha$, and $\beta$ constant over time.  Again these methods differ from ours as they do not allow for new topics, the key 
aspect of our anomalous pattern detection framework. SATM~\cite{SATM} extends topic models to use a temporal ordering of topics by learning a random topic 
initialization at $t_0$ and then allowing each time step to be based 
upon the previous slice similar to \cite{dDTM}.  Finally, the multiscale 
dynamic topic model~\cite{multiscale} allows $\phi$, the distribution of 
words in a topic, to vary over time slices. Here, topics are influenced 
through the Dirichlet hyperparameter which is adapted from a weighted 
sum of the empirical distributions of words at different time scales.  
The unifying factor in time variant topic models is that topics 
gradually change over time and smoothly evolve. Little work has been 
done on detecting newly emerging topics, which is the primary focus of our novel topic modeling approach.

Labeled LDA \cite{ramage2009labeled} is a closely related supervised topic model which can be adapted to the task of emerging topic detection by enforcing the constraint that background documents can contain only a subset of all topics known as background topics, while the foreground documents are allowed to contain background as well as foreground topics. Labeled LDA does not attempt to learn foreground topics that contrast with the background topics. Our experimental results below demonstrate that our proposed contrastive topic modeling approach much more accurately captures an emerging topic of interest as compared to Labeled LDA.

Our work is also related to the ``burst'' modeling approach in \cite{he:2010}. However, modeling term bursts cannot effectively detect a new event when it appears as a new co-occurrence of already popular terms affecting a small set of locations.

Topic models have also been extended to incorporate geographical information, and these methods have been applied to social media data for 
two primary purposes: 1) investigating regional variations in trending topics, language use, etc., and 2) predicting geographic locations of 
documents.  Eisenstein et al.~\cite{eisenstein2010latent} extend topic models by introducing a latent ``region'' variable into 
the graphical model so that $K$ topics are learned for each region. Yin et al.~\cite{yin2011geographical} focus on how to compare the identified topics across different regions, and Hong et al.~\cite{hong2012discovering} 
show a computationally efficient way to represent both users and geographical areas using sparse modeling techniques.  In general, these 
methods focus on training predictive models that are specific to individual areas but do not change over time, as opposed to our work which 
focuses on detecting emerging changes in an area.  We note that the subset of locations affected by an emerging event may not correspond 
well to the partitions learned from background data when no events are occurring, and thus the latent region variable may not effectively 
capture events of interest.

\section{Methodology}
Semantic Scan (SS) integrates a novel contrastive LDA model with spatial scan statistics to incorporate unstructured text data into a spatial event detection framework. The key assumption in doing so is that a novel event of interest will generate text documents which are similar to each other, yet different from the remainder of the corpus, in their co-occurrences of terms; note that the individual terms may appear elsewhere in the corpus in other contexts.  This can be thought of as a noisy-channel model where the true data stream of interest 
(consisting of all and only those documents corresponding to the novel event) has been obfuscated through the use of natural language such that it no longer exhibits an explicit, observable, expert label. However, under the assumption above, the labels can be approximately recovered through the use of a novel contrastive topic modeling approach, which we describe shortly. For example, in disease surveillance, each topic represents a class of diseases with similar symptoms. Each disease case is described by a patient to a health care provider and transcribed, introducing errors, abbreviations, and other variability, but the patterns of keywords in these descriptions can be used to group cases and thus provide useful structure for detection of anomalous trends.  Once the data is structured, spatial scan methods represent the state-of-the-art in terms of incorporating spatial and temporal data to identify emerging space-time patterns, which is important since we expect the event of interest to be localized in space and time as well as generating text data which forms a novel and coherent topic.

Thus our general Semantic Scan framework consists of three main steps: (i) Two sets of topics - background topics $\phi_k, k \in \{1 \ldots K\}$, and foreground topics $\phi'_{k'}, k' \in \{1 \ldots K'\}$, are learned from the data. Each topic represents a sparse probability distribution over words in the vocabulary. (ii) We perform online assignment of each individual document to the most likely topic, using a robust assignment method similar to expectation maximization (EM). (iii) We perform spatial scanning, identifying spatial regions that have a significantly higher than expected number of recent cases assigned to some foreground topic. We now provide details on the contrastive topic model used to learn topics within the Semantic Scan framework, and then describe our methods for online topic assignment and spatial scanning.

\subsection{Contrastive Topic Model}

We now describe our novel contrastive topic model, which builds on Latent Dirichlet Allocation~\cite{blei2003latent} but is specifically designed to detect anomalous, newly emerging topics. We note that while the last step of the Semantic Scan framework, spatial scanning, incorporates both spatial and temporal information, the initial topic modeling step does not use the spatial information, fitting the topic model using documents from all locations for a given time frame. Given that it is unknown whether an event is occurring, and that the subset of locations affected is uncertain, incorporating spatial information is difficult in this setting. Learning individual topic models for each spatial location suffers from both increased computation time and data sparsity issues, overfitting topics to underlying noise using a very small number of documents. Similarly, learning topic models for each potentially affected subset is computationally infeasible, thus motivating our proposed approach. Also, since exact inference in the LDA model is intractable, our LDA implementation uses a collapsed Gibbs sampler to perform approximate inference as in~\cite{gibbs}.  In addition to computational efficiency, this approach extendes more easily to the contrastive topic model described below, as compared to alternatives such as variational inference.

\begin{figure}[t]
\begin{center}
\includegraphics[width=5.5cm]{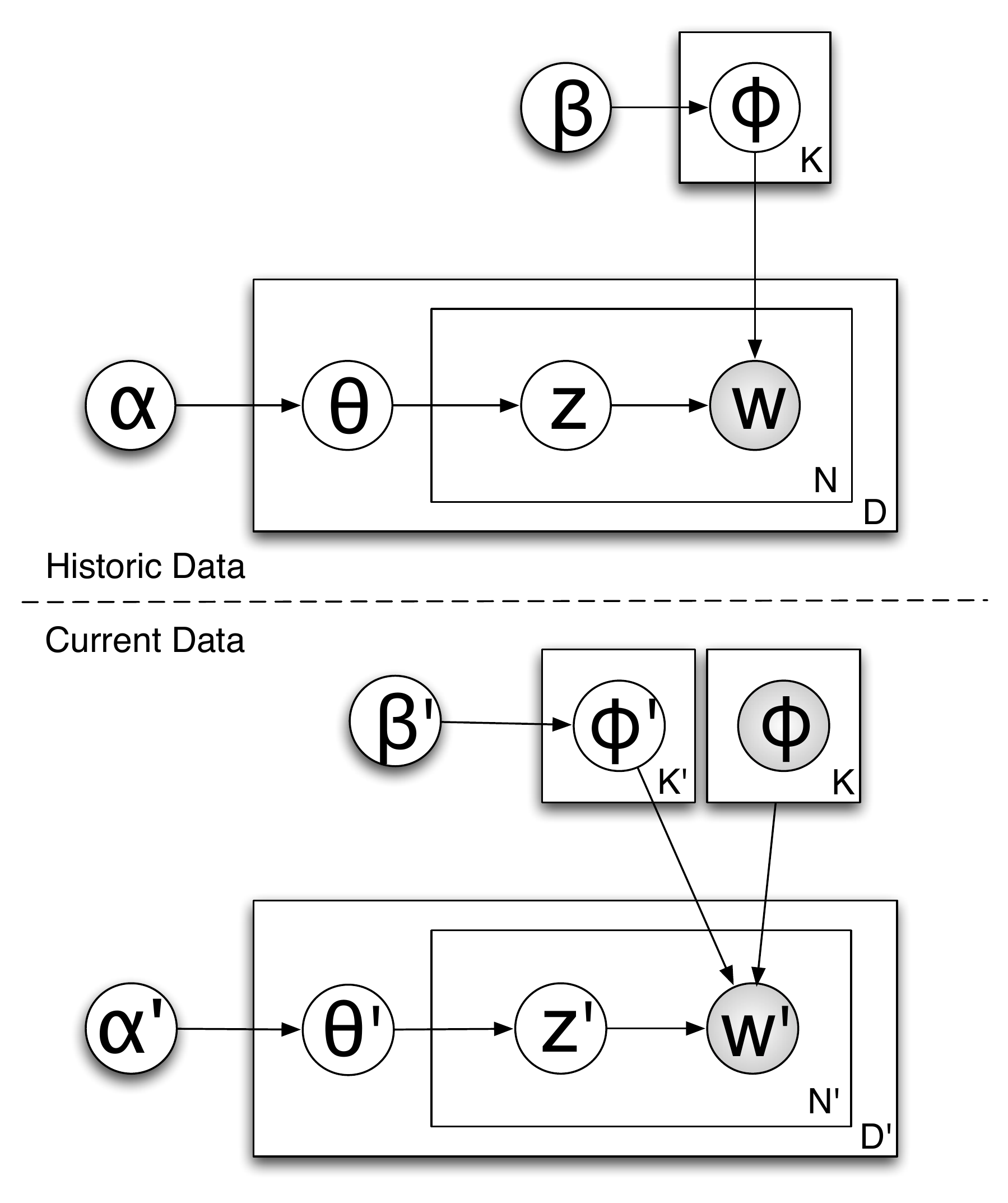}
\caption{Semantic Scan Topic Model.  A set of $K$ topics are learned using historical data and $K'$ topics are learned using current data.  Semantic Scan then re-fits the $K'$ foreground topics using the combined set of $K+K'$ topics, where $\phi_k$ are fixed for all background topics, treating these as observed variables.}\vspace{-0.3cm}
\label{model}
\end{center}
\end{figure}

Contrastive Topic Modeling consists of three steps: (i) In the background phase, it first learns a set of $K$ background topics using the corpus of historical training data. Once learned, these topics can be reused across many days of detection until a new emerging event is detected. (ii) In the first step of detection phase, it considers a moving window of $X$ days ($X=3$ in our implementations) and learns a separate set of $K'$ foreground topics using only documents from the moving window, using a standard LDA topic model. (iii) In the second step of detection phase, it now considers the combined set of all $(K+K')$ topics and refits the model as shown in the bottom panel of Figure~\ref{model}.
When the $K'$ topics are first learned, they can overlap significantly with the $K$ background topics, because the foreground documents in the moving window contain both background topics $\phi$ and foreground topics $\phi'$. By refitting the model $\phi'$ after introducing the fixed background topics $\phi$, the foreground topics $\phi'$ are forced to align with emerging topics in the foreground documents. This is because the background component of foreground documents is explained by and attributed to the background topics in the modified Gibbs sampling procedure (Algorithm~\ref{newgibbsSampler}).

The key difference is that the $K$ background topics are treated as fixed for this inference step, and thus the distribution over words 
$\phi_k$ for each background topic is treated as an observed variable in the graphical model.  The distribution over words $\phi'_{k'}$ for 
each of the $K'$ foreground topics is allowed to vary, as is the distribution $\theta_d$ over the $K+K'$ topics for each foreground document $d$. 
This has the effect of pushing the foreground topics toward capturing distributions over words in the current data that are not well 
modeled by the fixed, background topics, thus allowing the contrastive topic model to learn a new set of topics that better model novel emerging
events.  While some of the re-fitted foreground topics may capture noise or other irrelevant patterns in the data, we expect that novel 
patterns of interest will also be captured, and then the spatial scanning step will distinguish ``signal'' from ``noise'' topics. Our 
experimental results, discussed in detail below, show that this approach produces topics which much more precisely capture novel 
events, increasing signal strength and therefore detection power, as compared to state-of-the-art competing approaches, which also include irrelevant terms and documents in the learned topics.

We note that the contrastive topic model with both background and foreground topics is no longer equivalent to LDA, as it contains fixed observable distributions of topics 
which impact the posterior inference. This required modification of the collapsed Gibbs sampler (Algorithm~\ref{newgibbsSampler}).  First, 
the Gibbs sampler is initialized by assigning a topic $z_{di} = k$ to each word $w_{di}$ of each document $d$.  Instead of drawing these 
initial topic assignments uniformly at random, we use the distributions over words $\phi_k$ that were learned for the $K$ background and $K'$ 
foreground topics in the first two steps of the contrastive topic modeling method described before.  We assign word $w_{di}$ to topic $k$ with probability proportional to 
$\phi_k^{(w_{di})}$, the component of topic $k$'s distribution over words for the given word $w_{di}$.  Based on this initial assignment of 
words to topics, we update the foreground topics, computing new values of $\phi_k^{(w)}$, the probability of each word in the vocabulary 
given topic $k$:
\vspace{-1mm}
\begin{equation}
\phi_k^{(w)} = \frac{n_k^{(w)}+\beta}{n_k^{(.)}+|V|\beta},\label{phieqn}
\end{equation}\vspace{-1mm}
\noindent
where $n_k^{(w)}$ is the number of assignments of term $w$ to topic $k$, $n_k^{(.)}$ is the number of assignments of all terms to topic $k$,
$\beta$ is the Dirichlet hyperparameter for $\phi$, and $|V|$ is vocabulary size.
Similarly, we compute new values of $\theta_d^{(k)}$, the probability of each (background or foreground) topic given document $d$:\vspace{-1mm}
\begin{equation}
\theta_d^{(k)} = \frac{n_d^{(k)}+\alpha}{n_d^{(.)}+(K+K')\alpha},\label{thetaeqn}
\end{equation}\vspace{-1mm}
\noindent
where $n_d^{(k)}$ is the number of assignments of words in document $d$ to topic $k$, $n_d^{(.)}$ is the number of assignments of words in 
document $d$ to all topics, $\alpha$ is the Dirichlet hyperparameter for $\theta$, and $K+K'$ is the number of topics.  Sampling proceeds as 
usual, except that the $\phi_k$ distributions for all background topics remain fixed throughout the inference process, and are not changed when 
sampling the topic assignments.

\begin{algorithm}[t]
 \For{each document $d$}{
 \For{each word $w_{di}$ in document $d$}{
   Assign topic $z_{di} = k$ with probability proportional to $\phi_k^{(w_{di})}$ from initial static and dynamic models.
  }
  Compute $\theta_d$ using equation~(\ref{thetaeqn})\;
 }
 \For{each foreground topic $k$}{
   Re-compute $\phi_k$ using equation~(\ref{phieqn})\;
 }
 \While{not Converged}{
  \For{each document $d$}{
   \For{each word $w_{di}$ in document $d$}{
    Remove current topic assignment, $z_{di} = k$\;
    Update $\theta_d$; if foreground topic, also update $\phi_k$\; 
    \For{each topic $k$}{
    Compute: $\mbox{Pr}(z_{di} = k) \propto \phi_k^{(w_{di})} \theta_d^{(k)}$\;
    }
    Sample a new topic assignment, $z_{di} = k$\;
    Update $\theta_d$; if foreground topic, also update $\phi_k$\;
   }
  }
 }
 \caption{Modified Gibbs Sampler\label{newgibbsSampler}}
\end{algorithm}

We note that the contrastive topic model is designed to be very different than other topic modeling methods with time-varying topics: our method 
biases the emerging foreground topics to be contrastively different from existing topics, under the assumption that such topics will be most useful for novel 
event detection.  As discussed before, previous methods instead aim for a smooth evolution of topics over time, capturing trends 
in the current set of topics rather than identifying entirely new topics.

\subsection{Online Topic Assignment}

Online inference for new documents in topic models is a non-trivial problem, and is of increased importance in the spatial event detection 
framework as the signal of interest can be lost due to poorly chosen topic assignments. From a surveillance perspective, the most 
interesting aspects of a dataset can have a very low probability of occurrence, and thus dimensionality reduction techniques such as 
LDA can drown out the subtle, spatially localized signal.  We discuss above how the overall LDA procedure can be modified to focus on 
infrequent, newly emerging patterns, but even once the set of topics is learned, this problem reoccurs when assigning documents to topics.
We note that the corpus of foreground documents on which the topic models are learned does not include all cases needed to compute counts and baselines for spatial scanning, so we cannot just use the $\theta_d$ distributions learned while performing topic modeling.

We examined common online assignment methods such as Gibbs sampling, but found that the combination of low term frequencies (for the 
novel terms of particular interest) and short document lengths caused the initial random assignments of words to topics to have a profound 
impact. In particular, for short documents, Gibbs sampling may assign identical or nearly identical documents to different topics rather 
than grouping all of these into the same topic, resulting in a diluted signal and lower detection power.  Other common methods, such as 
summing the probabilities for a given topic over all words in the document, and then taking the maximum over topics, also did not perform 
well.  Summing probabilities tends to emphasize commonly occurring words (which may have high probabilities for multiple topics) even when 
these are not especially relevant.  On the other hand, multiplying probabilities tends to emphasize very rare words with small 
probabilities, and tends to be unduly influenced by misspellings and other rare but irrelevant words.  Thus we developed a method inspired 
by expectation-maximization (Algorithm~\ref{online}), and used this method to compute the assignment of documents to topics.

\begin{algorithm}[ht]
 $\theta_d^{(1)} = \dotsc = \theta_d^{(K)} = \frac{1}{K}$\;
 
 \While{not Converged}{
  \For{each word $w_{di}$ in document $d$}{
   \For{each topic $k$}{
    Compute: $\mbox{Pr}(z_{di} = k) \propto \phi_k^{(w_{di})} \theta_d^{(k)}$\;
   }
   Normalize: $\sum_k \mbox{Pr}(z_{di} = k) = 1$\;
  }
  
  \For{each topic $k$}{
   Compute: $\theta_d^{(k)} \propto \alpha + \sum_i \mbox{Pr}(z_{di} = k)$\;  
  }
  Normalize: $\sum_k \theta_d^{(k)} = 1$\;
 }
 
 Assign document $d$ to topic $k = \arg\max_k \theta_d^{(k)}$\;
 
 \caption{Online Document Assignment\label{online}}
\end{algorithm}

Given the distributions over words $\phi_k$ for each topic $k$, this algorithm assigns entire documents to topics, and can be 
performed independently for each document.  In Algorithm \ref{online}, $\phi_k^{(w_{di})}$ represents the probability of the $i^{th}$ word of 
document $d$ under topic $k$, and $\theta_d^{(k)}$ is the proportion of topic $k$ in the topic mixture for document $d$. $\theta_d$ is 
initialized uniformly for each document since that document may not have been seen previously and no prior knowledge is assumed. Also, we 
note that this assignment of documents to topics is performed after we have learned the distribution over words $\phi_k$ for each topic $k$.  
Unlike other online algorithms, we do not allow the topics to vary during the assignment phase: methods that allow topics to vary have a 
smoothing effect which can drown out the signal that we are trying to find. Also, topic assignment is deterministic for each document, 
rather than dependent on random initializations, so identical documents will always be assigned to the same topic.  This is important to avoid diluting the signal of interest in the spatial scan step below.  
We use this online document assignment approach to classify all documents in the moving detection window ($W_{max} = 3$ days for our experiments) and the preceding 30 days, which are used to estimate the expected counts used in the spatial scanning step below.  We note that each document is assigned to either a foreground or background topic, but we perform the spatial scan over foreground topics only, thus ignoring any document assigned to a background topic.

\subsection{Spatial Scanning}

Once each document has been assigned to one of the $K+K'$ topics, we can perform a spatial scan by first computing the aggregate count (number of 
documents assigned to that topic) $c_{i,k}^t$ for each spatial location (zipcode) $s_i$ for each foreground topic $k$ for each day $t$.  We then compute 
the expected counts i.e. baselines $b_{i,k}^t$ for each location, foreground topic, and day, using a 30-day moving average.  Finally, for each foreground topic $k$, 
we scan over all spatio-temporal regions $S = S_{spatial} \times S_{temporal}$, where $S_{spatial}$ is a circular spatial region consisting 
of some center location $s_i$ and its $n$-nearest neighbors in Euclidean distance (for all locations $s_i$ and all $n \in \{1 \ldots 
n_{max}\}$), and $S_{temporal}$ is a temporal window consisting of the most recent $W$ days (for all $W \in \{1 \ldots W_{max}\}$).  For 
each such spatio-temporal region $S$, with corresponding aggregate count $C(S)$ and aggregate baseline $B(S)$, we compute the log-likelihood ratio $F(S)$ using the 
expectation-based Poisson scan statistic defined in equation~(\ref{ebpequation}). We evaluate all combinations of space-time region and topic, and return the highest scoring space-time region $S$ with its associated score $F(S)$ and topic $k$. There are $N \cdot n_{max}$ circular spatial regions to consider, where $N$ is the number of locations and $n_{max}$ is the maximum neighborhood size.  For each spatial region, we must consider $W_{max}$ time durations and $K'$ foreground topics, for a total complexity of $O(N \cdot n_{max} \cdot W_{max} \cdot K')$ for the spatial scan step. In our ED experiments and Yelp experiments below, we had $N=97$ and $N=58$ locations respectively, and we used $K=25$ background topics and $K'=25$ foreground topics for both experiments.  Randomization testing can be used to test for statistically significant clusters, correctly adjusting for multiple hypothesis testing, as in~\cite{kulldorff}.

\section{Experiments}

\begin{figure*}[t]
\centering
\begin{tabular}{c c c}
  \centering
  \includegraphics[page=1,trim=28mm 45mm 35mm 35mm,clip=true,width=0.3\linewidth]{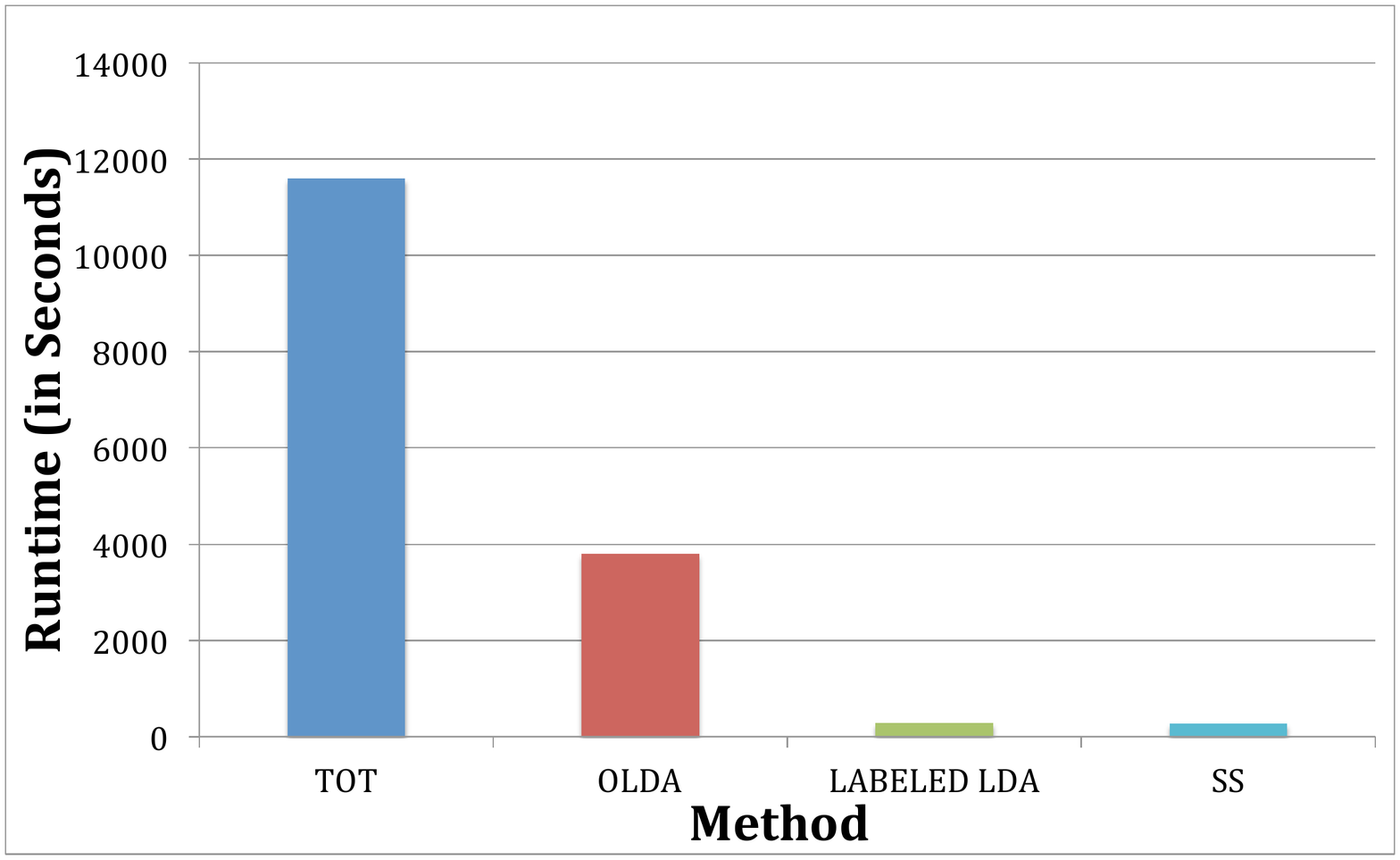} & \includegraphics[page=5,trim=28mm 45mm 30mm 35mm,clip=true,width=0.3\linewidth]{images/graphs_ed.pdf} & \includegraphics[page=6,trim=28mm 45mm 30mm 35mm,clip=true,width=0.3\linewidth]{images/graphs_ed.pdf} \\
  (a) Runtimes (ED) & (b) Fraction of Outbreaks Detected (ED) & (c) Days to Detect (ED)  \\
  & & \\
  \includegraphics[page=1,trim=28mm 45mm 35mm 35mm,clip=true,width=0.3\linewidth]{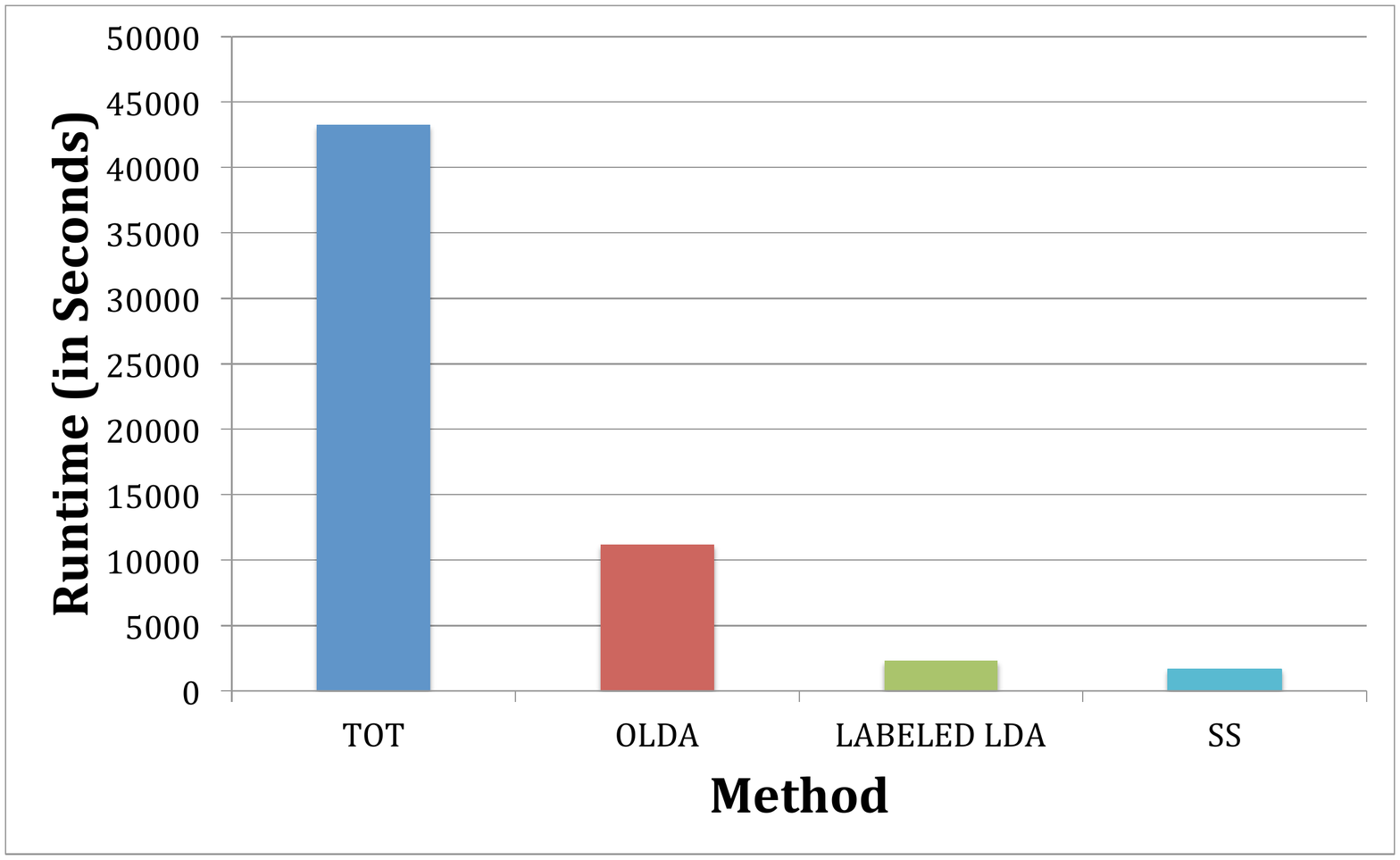} & \includegraphics[page=5,trim=28mm 45mm 30mm 35mm,clip=true,width=0.3\linewidth]{images/graphs_yelp.pdf} & \includegraphics[page=6,trim=28mm 45mm 30mm 35mm,clip=true,width=0.3\linewidth]{images/graphs_yelp.pdf} \\
  (d) Runtimes (Yelp) & (e) Fraction of Events Detected (Yelp) & (f) Days to Detect (Yelp)  \\
  & & \\
\end{tabular}
\vspace{-3mm}
\caption{Figures (a) and (d) on the left show the runtimes of SS compared to various methods on the ED and Yelp datasets. Figures (b) and (e) show the fraction of events detected as a function of number of false positives per year. Figures (c) and (f) show the average number of days required to detect after start of the outbreak as a function of false positives per year.}
\label{fig:fig1}
\end{figure*}

\begin{figure*}[t]
\centering
\begin{tabular}{c c c}
  \centering
  \includegraphics[page=2,trim=28mm 45mm 30mm 35mm,clip=true,width=0.3\linewidth]{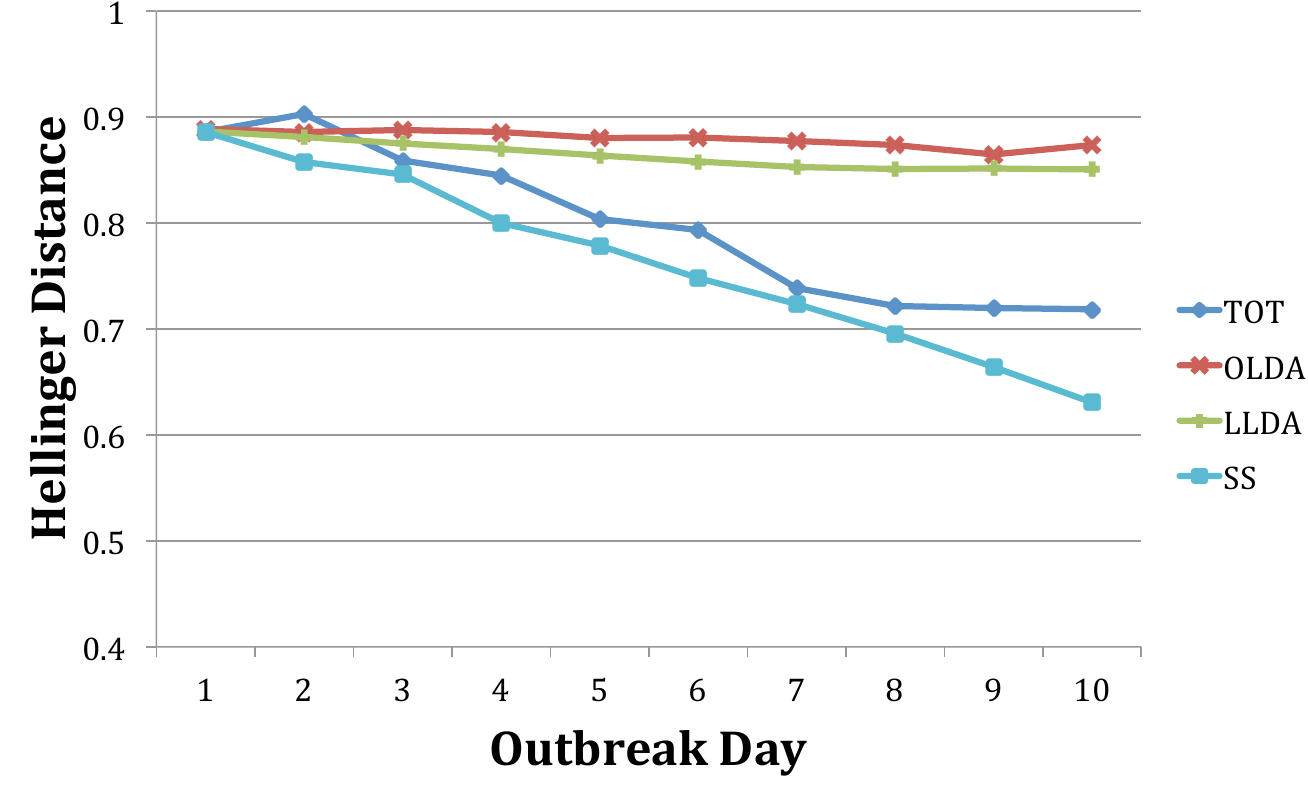} & \includegraphics[page=4,trim=28mm 45mm 30mm 35mm,clip=true,width=0.3\linewidth]{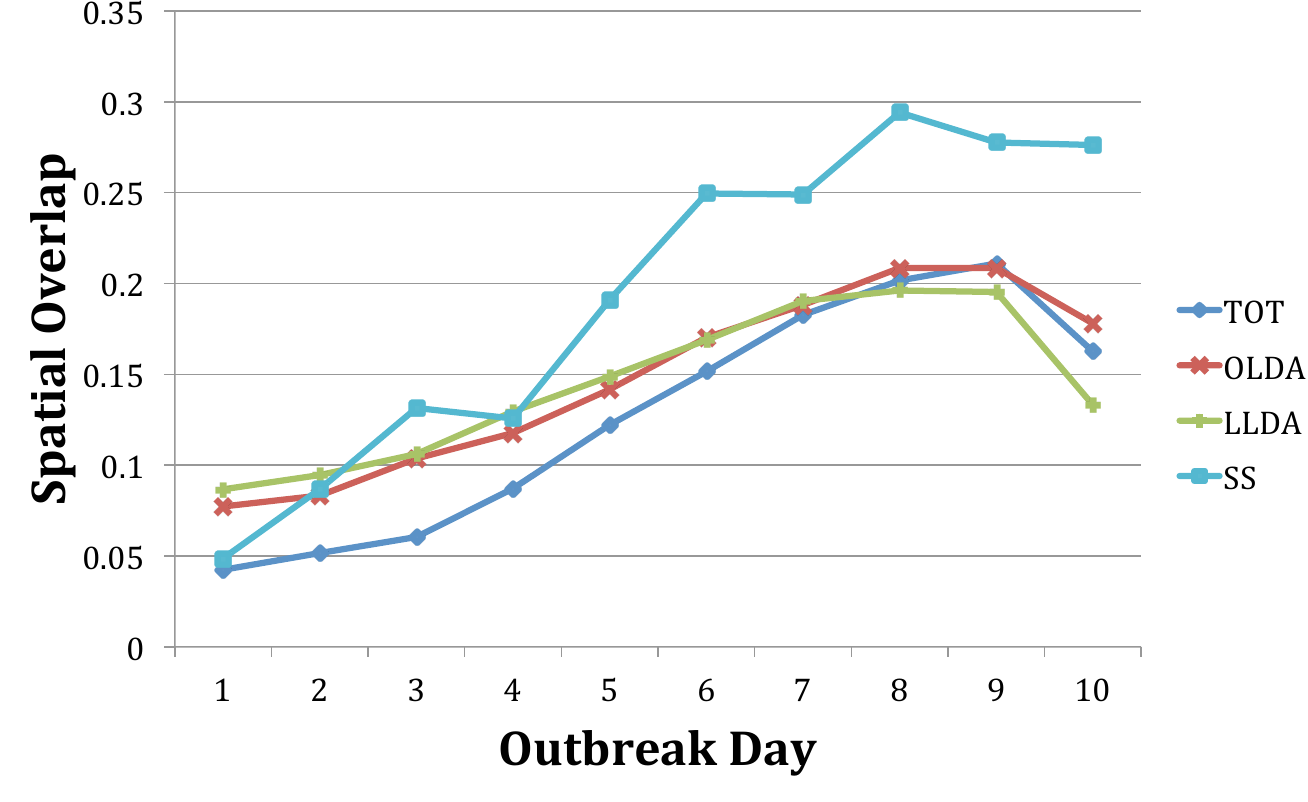} & \includegraphics[page=3,trim=28mm 45mm 30mm 35mm,clip=true,width=0.3\linewidth]{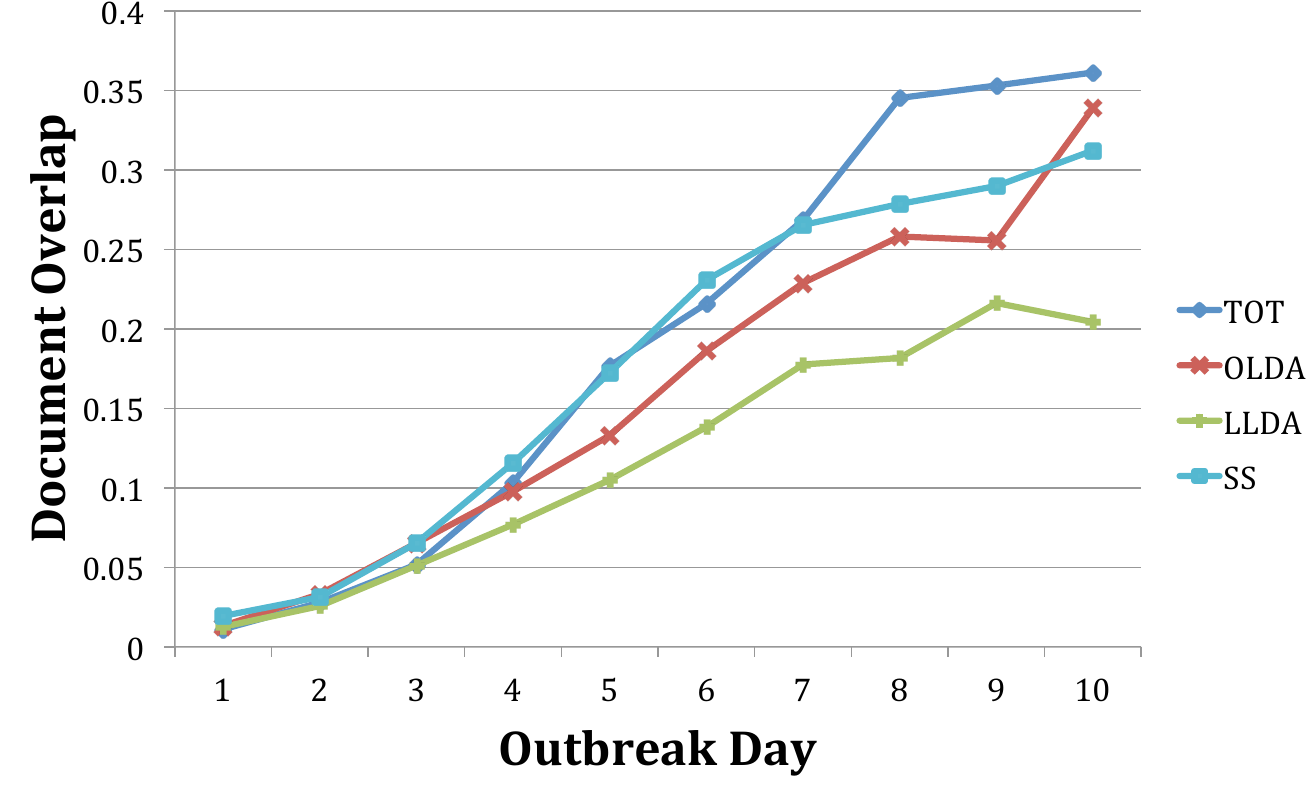} \\
  (a) Hellinger Distances (ED) & (b) Spatial Overlap (ED) & (c) Document Overlap (ED) \\
  & & \\
  \includegraphics[page=2,trim=28mm 45mm 30mm 35mm,clip=true,width=0.3\linewidth]{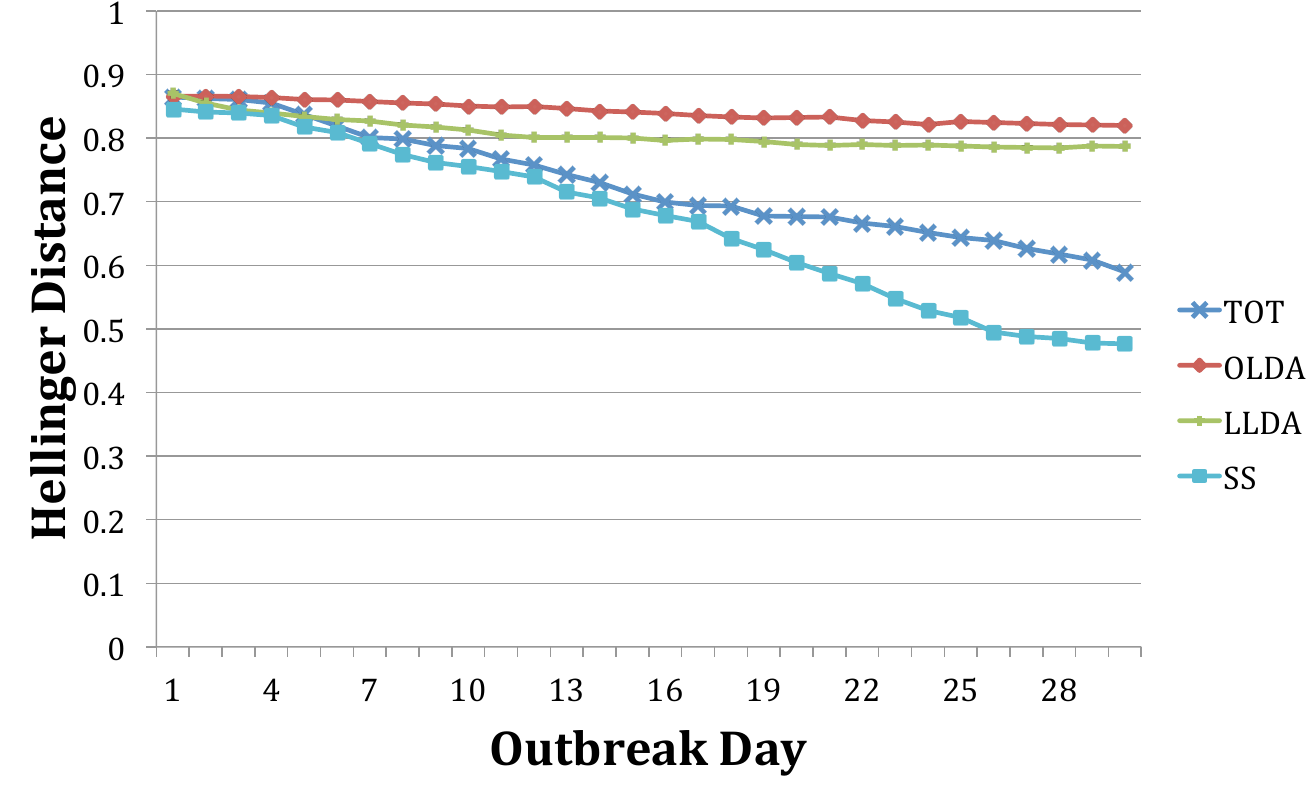} & \includegraphics[page=4,trim=28mm 45mm 30mm 35mm,clip=true,width=0.3\linewidth]{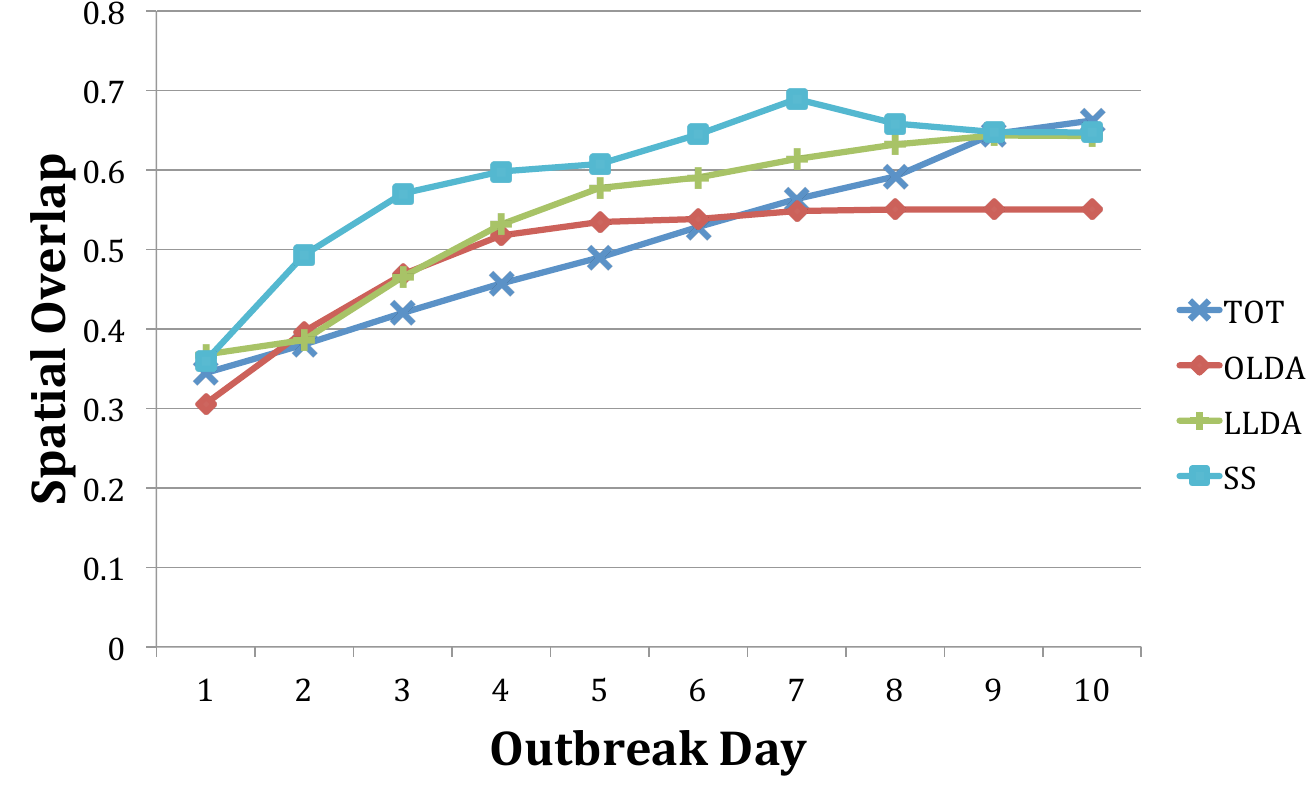} & \includegraphics[page=3,trim=28mm 45mm 30mm 35mm,clip=true,width=0.3\linewidth]{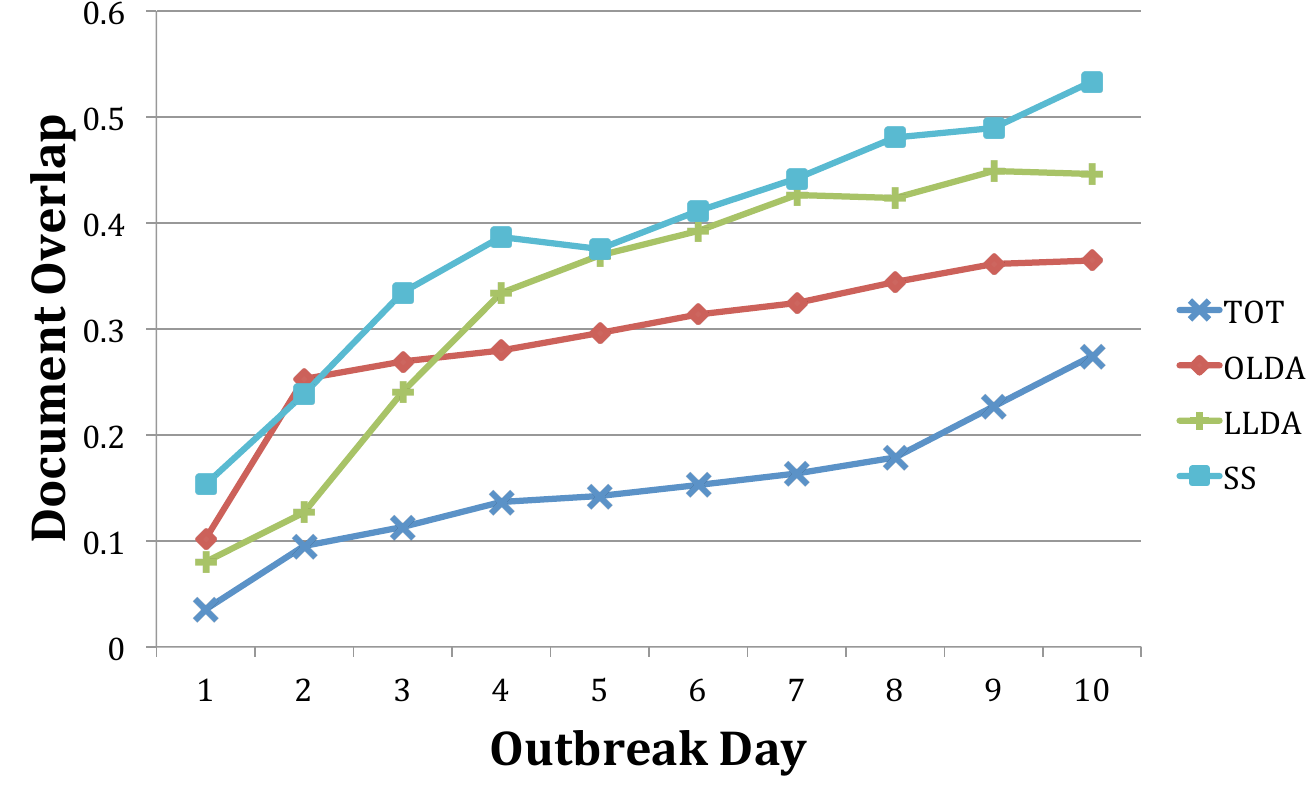} \\
  (d) Hellinger Distances (Yelp) & (e) Spatial Overlap (Yelp) & (f) Document Overlap (Yelp) \\
  & & \\
\end{tabular}
\vspace{-3mm}
\caption{Figures showing the effectiveness of the three stages of detection. The top figures show the average Hellinger distance, spatial overlap and document overlap as function of outbreak day on the ED dataset. The figures in the bottom row show the same metrics on the Yelp dataset.}
\label{fig:fig2}
\end{figure*}

\begin{figure*}[t]
\centering
\begin{tabular}{c c}
  \centering
  \includegraphics[page=6,trim=45mm 30mm 50mm 45mm,clip=true,width=0.5\linewidth]{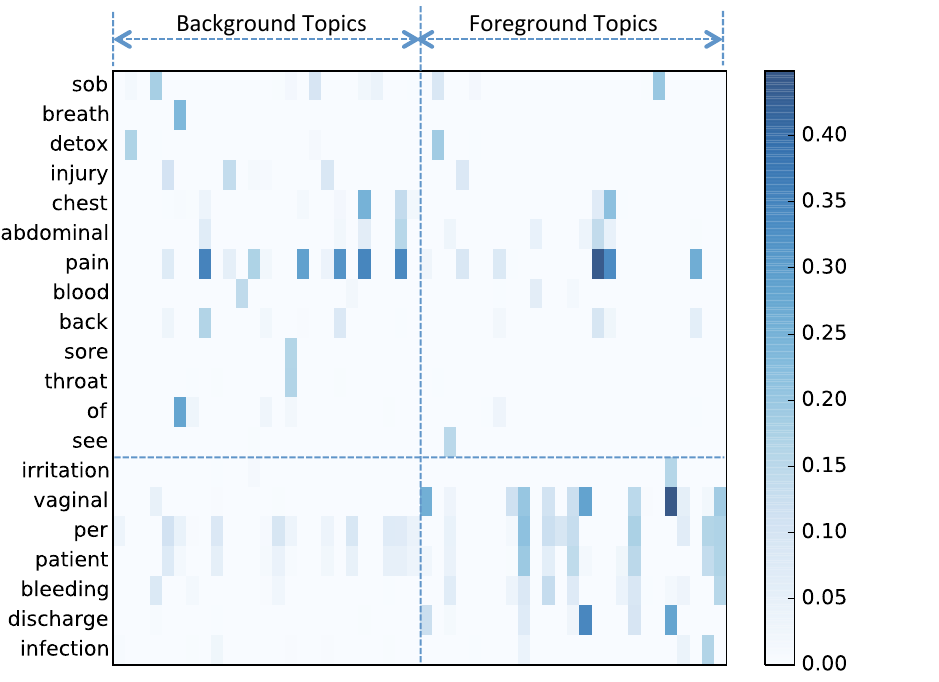} & \includegraphics[page=7,trim=45mm 30mm 50mm 45mm,clip=true,width=0.5\linewidth]{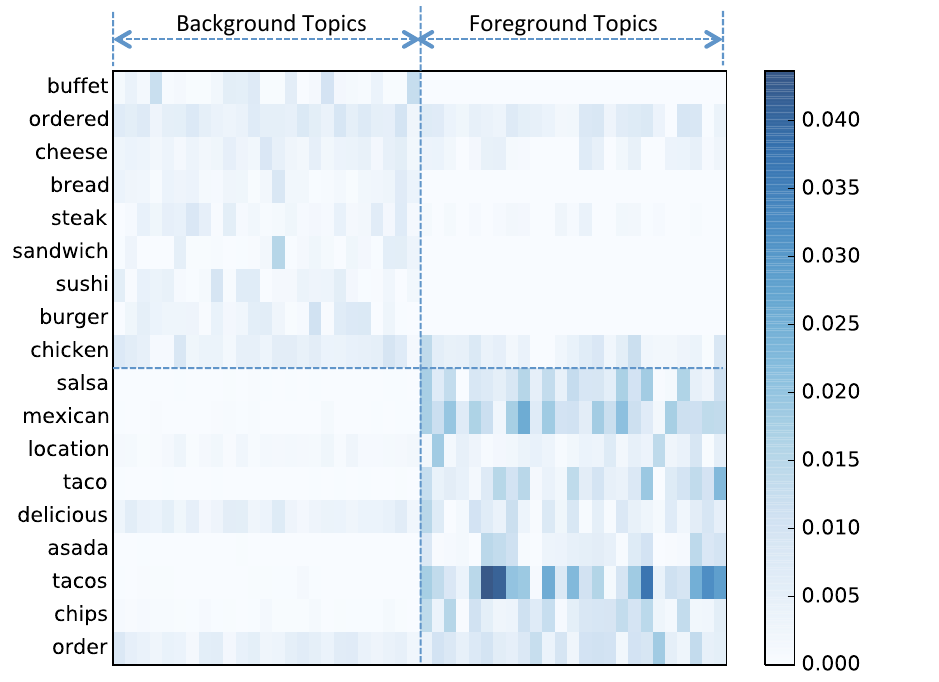} \\
  (a) ED Topics & (b) Yelp Topics \\
  & \\
\end{tabular}
\vspace{-3mm}
\caption{Example background and foreground topics detected from the ED and Yelp datasets. Subfigure (a) shows topics learned on the ED dataset during an outbreak of a sexually transmitted disease, while subfigure (b) shows topics on the Yelp dataset characterizing the emergence of Mexican restaurants in a Las Vegas neighborhood. For each dataset, we show the top words from both the background and foreground topics. Words above the horizontal blue line are top words with significant presence in the background topics, while words below the line are dominant in the foreground topics.}
\label{fig:fig3}
\end{figure*}

We compared Semantic Scan to three alternative state-of-the-art methods such as Topics over Time, Online LDA, and Labeled LDA on two real-world tasks: (i) a disease surveillance task monitoring free-text Emergency Department chief complaints in Allegheny County, and (ii) an emerging business trend detection task based on Yelp reviews. The four methods we compared are as follows:

\begin{itemize}
  \item {\bf Semantic Scan (SS)}: We used a 3 day moving window for detection, $K=25$ background topics, and $K'=25$ foreground topics.  Typical hyperparameter values $\alpha = \frac{1}{K+K'}$ and $\beta = \frac{1}{|V|}$ were used, where $|V|$ is the vocabulary size.

  \item {\bf Topics over Time (ToT) \cite{wang2006topics}}: Since SS learns 50 total topics ($K=25,K'=25$), detects events with a 3-day moving window and uses the past 30 days to calculate expected counts for spatial scanning, we ran a comparable detection with ToT by using a moving window of 33 days and learning 50 topics in each window, using hyperparameters as described in \cite{wang2006topics}.

  \item {\bf Online LDA (OLDA) \cite{hoffman2010online}}: We ran OLDA with a similar detection window of 33 days, learning 50 topics in each window. We set the OLDA hyperparameter $\kappa$, which controls the rate at which topics being learned in an online fashion are updated, to 0.9. 

  \item {\bf Labeled LDA (LLDA) \cite{ramage2009labeled}}: In order to provide appropriate supervision to Labeled LDA, we assigned labels $1, 2 \ldots, K$ to background documents and $1, 2 \dots, K+K'$ to foreground documents. This implies that we constrained background documents to contain only the $K=25$ background topics, and allowed foreground documents to contain both the $K=25$ background as well as the $K'=25$ foreground topics.
\end{itemize}

We implemented SS, ToT, and Labeled LDA in Python, and used the publicly available Python implementation of OLDA in our comparisons.

After the topics are learned, the online document assignment and spatial scanning steps for the competing approaches are identical to SS. The first 30 days of the window were used for calculating expected counts, while the last 3 days were used for event detection.  We note that a circular spatial scan~\cite{kulldorff}, using a 30-day moving average to estimate expected counts and a maximum neighborhood size $n_{max} = 30$, was used for all methods.  We expect the circular scan to have high detection power for compact spatial clusters, while alternative approaches such as the fast subset scan~\cite{LTSS} would have higher power to detect highly elongated or irregular clusters.  Similarly, more complex time series analysis methods could be incorporated to account for seasonal and day of week trends, time-varying covariates, etc.  These alternative scan approaches and time series analysis methods could be easily plugged into the semantic scan framework.

\subsection{ED Dataset}

The first dataset we use is a de-identified spatio-temporal dataset of hospital Emergency Department (ED) data collected from 
ten Allegheny County, Pennsylvania hospitals from January 1, 2003 to December 31, 2005. The dataset consisted of $\sim$340K records of individual ED visits, each of which contained four attributes: admission date, home zipcode, chief complaint, and International Classification of Diseases-9th Edition (ICD-9) code.  They appear similar to the records given in Table~\ref{cases}. The first three attributes are populated upon a patient's admittance to the ED, while ICD-9 code is generally not populated until the patient's discharge. We use this attributes for evaluation and comparison purposes only.

\begin{table}[!ht]
\caption{Example Case Formats}
\label{cases}
\centering
\small
\tabcolsep=0.11cm
\begin{tabular}{|c|c|l|c|}
\hline
Date & Location & Chief Complaint & ICD-9 \\
\hline
01.01.2004 & 15213 & COUGH/NAUSEA & 789 \\
02.03.2004 & 15232 & BLEEDING & 556.3 \\
07.04.2005 & 15232 & ETOH & 421 \\
\hline
\end{tabular}
\end{table}

The ``Chief Complaint'' field of the dataset is a free-text field recorded by a triage nurse upon a patient's admittance to the ED. Chief complaints are generally short (a few words or a phrase, such as ``pain in rt arm''), have little grammatical structure and minimal standardization across records, and are very noisy, with frequent misspellings and inconsistent use of terms and abbreviations.  The entries in the ``ICD-9'' field of the dataset are standardized codes used to \emph{manually} classify diseases and ailments into specific groups. They are primarily used for billing purposes in the United States; we use these for our leave-one-out evaluation strategy.

We split the dataset into the background documents from 2003 which are used to learn background topics, and foreground documents from 2004-2005 which are used for the detection. We created 10 event outbreaks for each of the 10 most frequent ICD-9 codes, i.e. 100 events in all. In order to create an outbreak corresponding to a given ICD-9 code, all complaints with that ICD-9 code were removed from the background and foreground documents, thus simulating the occurrence of a novel, previously unseen outbreak in the data.  We then used the Bayesian Aerosol Release Detector (BARD), a publicly available outbreak simulator~\cite{hogan07}, to inject a disease outbreak into the foreground documents.  The BARD simulator is a highly realistic model of the spatio-temporal distribution of Emergency Department cases resulting from a bioterrorist attack (airborne release of weaponized anthrax spores), integrating a dispersion model which takes wind speed and direction into account with a detailed, population-based patient model to predict who will be affected and when they will visit a hospital Emergency Department.  The text complaints injected as part of the outbreak were uniformly sampled from the held-out complaints for the given ICD-9 code, thus modeling a novel outbreak which affects the population similarly to anthrax but may have very different symptoms.

Minimal preprocessing was done to the chief complaint data field: all words were converted to lowercase, punctuation was removed, and 
tokens with slashes, ampersands, and similar punctuation were separated into two words. In many applications, text normalization 
methods such as stop word lists and stemming are used to preprocess data, but we did not do this as these methods could negatively 
impact detection power by removing the signals of interest in this domain. No methods were provided access to the ICD-9 field, which was used for evaluation purposes only. 

\subsection{ED Results}

First, we compare the runtimes of the various detection algorithms. Figure~(\ref{fig:fig1}.a) shows average runtimes of the various methods per injected outbreak, i.e., runtime for a single detection run through the entire dataset. The average runtime for SS is 270.8 seconds, much faster than
ToT and OLDA (11595 and 3786 seconds respectively) and slightly faster than Labeled LDA (280.8 seconds).

Second, we evaluate timeliness of detection of the various methods. To evaluate the performance of SS compared to other methods for detecting disease outbreaks, we plot the average number of days taken to detect an outbreak and the fraction of outbreaks detected as a function of the allowable false positive rate in figures (\ref{fig:fig1}.b) and (\ref{fig:fig1}.c) respectively. For a fixed false positive rate of 1/month, a level typically considered acceptable by public health practitioners, it takes 4.97 days on average for SS to detect an outbreak, versus 6.05, 6.44, and 6.34 for ToT, OLDA, and Labeled LDA respectively. For the same fixed false positive rate, the percentage of outbreaks correctly detected was 92\% compared to 81\%, 77\%, 77\% for ToT, OLDA, and Labeled LDA respectively.

Third, we evaluate the ability of Semantic Scan to precisely characterize novel outbreaks. This was measured by computing the average Hellinger distance (HD) between the true distribution of injected cases and the point estimate of the distribution over words $\phi_k$ for the detected topic, as a function of the number of days since the start of the outbreak. From figure~(\ref{fig:fig2}.a), we see that SS has the lowest HD of all methods. Therefore, we conclude that the distribution $\phi_k$ of the detected topic was closest to the true outbreak distribution for SS, implying that this approach is better at fitting a topic to novel emerging trends.

Fourth, we compare the accuracy of the identified spatial outbreak regions, measuring the Jaccard overlap between the set of detected locations and the set of true affected locations, which we call Spatial Overlap (SO).  As shown in figure~(\ref{fig:fig2}.b), we observe that SS outperforms other methods as the outbreak days progress.

Finally, we compare the accuracy of the online document assignment step by measuring the Jaccard overlap between the set of documents detected to contain the emerging topic the set of documents which were truly generated from a mixture containing the emerging topic, which we call Document Overlap (DO). From figure~(\ref{fig:fig2}.c), we observe that SS has the best document overlap, comparable only to ToT which has a much higher runtime.

In figure (\ref{fig:fig3}.a), we show an example of the topics learned on the ED dataset. The 25 background topics were learned from 2003 data, while the 25 foreground topics were learned from a simulated outbreak of ICD-9 code 623 corresponding to sexually transmitted diseases (STDs). The figure shows the top words based on the maximum probability of their occurrence in the learned topics. Words above the horizontal blue line in the figure such as \emph{pain}, \emph{throat}, \emph{injury}, etc. correspond to significant words in the background topics, while those below the line such as \emph{vaginal}, \emph{irritation}, \emph{infection}, etc. have significant presence in the foreground topics. Some of the foreground topics describe the disease outbreak very strongly. Words dominant in the foreground topics such as \emph{vaginal}, \emph{bleeding}, \emph{discharge}, \emph{infection}, etc. accurately characterize the ICD9 code 623. While these words were also present in the background documents, their strong co-occurrence in the foreground documents allowed us to detect and characterize the disease outbreak.



Overall, we can see that for novel emerging outbreaks, SS achieves much more timely and accurate detection than the competing approaches.

\subsection{Yelp Dataset}

Our second dataset is the Yelp reviews dataset that was publicly released by Yelp for academic research purposes\footnote{Available at \url{http://www.yelp.com/dataset_challenge}}. We limited ourselves to businesses and reviews from Las Vegas only. Each review was considered as a document. Its location was associated to the zipcode of the business it belonged to. Reviews before Jan 1, 2014 were considered as background documents, and reviews after this date formed the set of foreground documents.

In order to create realistic events depicting emerging business trends, we use the categories that Yelp businesses are associated with. In particular, we choose 70 categories of restaurants such as Greek, Bangladeshi, Croatian, Thai, Burmese, etc. For creating a simulated emerging business event to test SS and competing methods, we pick a particular category of restaurants and remove all reviews corresponding to this category of restaurants from the background and foreground documents. We then introduce an event corresponding to surge in a particular type of restaurants in a city neighborhood by introducing the heldout reviews for that restaurant type into the foreground corpus. Each introduced event was assumed to last thirty days, with an expected $20*d$ reviews injected into the data on the $d^{th}$ day of the simulated event, and affected a circular region consisting of 30 zipcodes whose center was randomly sampled from the set of zipcodes in Las Vegas.

The preprocessing steps to clean the business reviews were identical to those applied to the ED dataset - non-alphanumeric characters were replaced with spaces, and all letters were converted to lower case. One major difference of this dataset from the ED dataset previously analyzed was the length of documents. The reviews for Yelp businesses are much longer than ED chief complaints, which is reflected in differences in the results, particularly in the days taken to detect an event.

\subsection{Yelp Results}

The metrics on which we compare the various methods to Semantic Scan are identical to those used in analysis of the ED dataset.

First, we compare the runtimes of the various detection algorithms. Figure~(\ref{fig:fig1}.d) shows average runtimes of the various methods per injected event. The average runtime for SS is 1649 seconds, much faster as compared to 43281, 11172, and 2320 seconds for ToT, OLDA, and Labeled LDA respectively.

Second, we evaluate timeliness of event detection of the various methods. To evaluate the performance of SS compared to other methods for detecting emerging business trends, we plot the average number of days taken to detect an event and the fraction of events detected as a function of the allowable false positive rate in figures (\ref{fig:fig1}.e) and (\ref{fig:fig1}.f) respectively. For a fixed false positive rate of 1/month, it takes 2.44 days for SS to detect an event, versus 5.71, 3.29, and 4.11 for ToT, OLDA, and Labeled LDA respectively. For the same fixed false positive rate, the percentage of outbreaks correctly detected was 97\% compared to 85\%, 80\%, 85\% for ToT, OLDA, and Labeled LDA respectively.

Third, we evaluate the ability of Semantic Scan to precisely characterize novel outbreaks. This was measured via the average Hellinger distance (HD) between the true distribution of injected cases and the point estimate for the detected topic as described before. From figure~(\ref{fig:fig2}.d), we see that SS has the lowest HD of all methods, and conclude that the topic detected by SS best characterized the emerging business event that was hidden in the data.

Fourth, we compare the accuracy of the identified spatial outbreak regions, measuring the Spatial Overlap between the set of detected locations and the set of true affected locations.  As shown in figure~(\ref{fig:fig2}.e), we observe that SS outperforms other methods especially in the early stages of the emerging event.

Finally, we compare the accuracy of the online document assignment step by measuring the Document Overlap between the set of documents detected to contain the emerging topic and the true set of injected documents. From figure~(\ref{fig:fig2}.f), we observe that SS has the best document overlap.

In figure (\ref{fig:fig3}.b), we show an example of the background and foreground topics learned on the Yelp dataset. The event in this case corresponded to a simulated increase in the number of Mexican restaurants in a Las Vegas neighborhood. Compared to ED complaints, Yelp reviews are much longer. While ED complaints tend to describe the entire case in typically less than 5 words, Yelp reviews span multiple sentences. As a result, the background topics tend to be more diffuse than the ED background topics. However, the foreground topics are similar to the ED foreground topics in that they tend to concentrate on fewer specific words. In this case, words such as \emph{tacos}, \emph{asada}, \emph{salsa}, \emph{chips}, etc. characterize the emerging surge of Mexican restaurants through their cooccurrence in the Yelp reviews. Similar to ED dataset, we observe a biclustering of words across the background and foreground topics as seen in figure (4.b).

\section{Conclusions}

In this paper, we presented Semantic Scan (SS), a novel framework for detecting anomalous patterns in spatio-temporal free text data. Through a comprehensive evaluation comparing SS with three state-of-the-art methods on two real-world tasks, we demonstrated that SS improves both detection and characterization of novel emerging events. We demonstrated how free text data can be incorporated into spatial event detection by integrating a novel contrastive topic model with robust online document assignment and spatial scanning, resulting in significant improvements in detection power and the characterization of the emerging event. In fact, as compared to many deployed systems, the improvement in timeliness of detection may be even greater in practice, given that free-text data is often available far earlier than the corresponding structured data (e.g., disease codes in the ED dataset, or business categories in the Yelp dataset).

Thus, SS is able to effectively detect and precisely characterize emerging events. SS discovers previously unseen anomalous textual patterns without any manual intervention, allowing users of the algorithm to formulate earlier and more targeted responses than current methods. Unstructured text information is far more pervasive than structured human annotation of events. As a result, event detection systems such as disease surveillance systems and online business review portals will have access to ever greater quantities of geotagged, unstructured text data, necessitating the need for novel methods for spatial event detection from free text. We therefore anticipate that Semantic Scan will be employed by a variety of users looking for a fast, scalable tool for unstructured event detection.


\section{Acknowledgments}
This work was partially supported by NSF grants IIS-0916345, IIS-0911032, and IIS-0953330. We wish to thank Alexandra Chouldechova for providing insightful criticism on this research project.

%
\bibliographystyle{unsrt}
\bibliography{semantic_scan}  
\end{document}